\documentclass[aps,prc,preprint,groupedaddress,superscriptaddress]{revtex4-1}
\usepackage[utf8]{inputenc}
\usepackage{graphicx}
\usepackage{amsmath, amssymb, bm, mathrsfs, gensymb, longtable}

\begin{document}

\title{Nuclear structure studies of double Gamow-Teller strength}

\author{N. Auerbach}
\affiliation{School of Physics and Astronomy, Tel Aviv University, Tel Aviv 
69978, Israel}
\author{Bui Minh Loc}
\email[]{minhlocbui@mail.tau.ac.il}
\affiliation{School of Physics and Astronomy, Tel Aviv University, Tel Aviv 
69978, Israel}

\date{\today}

\begin{abstract}
The double Gamow-Teller strength distributions in even-$A$ Calcium isotopes 
were calculated using the nuclear shell model by applying the single 
Gamow-Teller operator two times sequentially on the ground state of the 
parent nucleus. The number of intermediate states actually contributing to the 
results was determined. The sum rules for the double Gamow-Teller operator in 
the full calculation were approximately fulfilled. In the case that the 
symmetry is restored approximately by introducing degeneracies of the 
$f$-levels, and the $p$-levels in the $fp$-model space, the agreement with the 
sum rules was very close.
\end{abstract}

\maketitle

\section{Introduction}
The double charge-exchange (DCX) processes are a promising tool to study nuclear 
structure in particular nucleon-nucleon correlations in nuclei. In the 1980s, 
the DCX reactions using pion beams that were produced in the three meson 
factories at LAMPF, TRIUMF, and SIN were performed.
Studies at lower pion energies ($E \leq 50$ MeV) have indeed produced clear 
signals of nucleon-nucleon correlations \cite{PhysRevLett.52.105, 
PhysRevLett.54.1482, PhysRevC.35.1425, PhysRevC.37.902, WEINFELD199033} and 
were successfully explained by the theoretical studies 
\cite{PhysRevLett.59.1076, PhysRevC.38.1277}. The pion DCX experiments 
discovered the double isobaric analog states (DIAS) 
\cite{PhysRevC.35.1334, PhysRevC.36.1479}.
At higher pion energies ($E > 300$ MeV), the studies discovered the giant dipole 
resonances (GDR) built on the IAS \cite{PhysRevLett.60.408, PhysRevC.38.2709, 
PhysRevC.40.850, PhysRevC.43.1111}, and double giant dipole resonances (DGDR) 
\cite{PhysRevLett.61.531, PhysRevC.41.202, PhysRevC.43.1318, PhysRevC.43.R1509} 
(See Refs. \cite{PhysRevLett.60.408, PhysRevLett.61.531} for definitions). 

At present, there is a renewed interest in DCX reactions, to a large extent due 
to the extensive studies of double beta-decay, both the decay in which 
two neutrinos are emitted ($2\nu\beta\beta$) and neutrinoless double beta 
($0\nu\beta\beta$) decay. In DCX and $\beta\beta$ decay, two nucleons are 
involved. The pion, however, interacts weakly with states involving the spin 
and the pion DCX reactions do not excite the states involving the spin, such as 
the double Gamow-Teller (DGT) state. The DGT strength is the essential part of 
the $\beta\beta$ decay transitions. It was suggested in the past that one 
could probe the DGT state and hopefully the $0\nu\beta\beta$ decay using DCX 
reactions with light ions \cite{ZHENG1990343, Cappuzzello2015}.

In present days, DCX reactions are performed using light ions. There is a 
large program called NUMEN in Catania where reactions with $^{18}$O, 
and $^{18}$Ne have been done \cite{Cappuzzello2018}. The hope is that such 
studies might shed some light on the nature of the nuclear matrix element of the 
double beta-decay and serve as a “calibration” for the size of this matrix 
element. These DCX studies might also provide new interesting information about 
nuclear structure. One of the outstanding resonances relevant to the 
$0\nu\beta\beta$ decay is the double Gamow-Teller (DGT) resonance suggested in 
the past \cite{AUERBACH198977, ZHENG1990343}. At RIKEN, there is a DCX program 
using ion beams for the purpose of observing DGT states and other nuclear 
structure studies \cite{UesakaCommu}. At Osaka University, the new DCX reactions 
with light ions were used to excite the double charge exchange state and compare 
to the pion DCX reaction results. One additional peak appeared in 
the cross-section suggesting that it is a DGT resonance \cite{Ejiri2017}. 

The DGT strength distributions in even-$A$ Neon isotopes was discussed in Ref. 
\cite{MUTO199213} and recently the calculation of DGT strength for 
$^{48}$Ca was performed in Ref.~\cite{PhysRevLett.120.142502}. 
In the present paper, the DGT transition strength distributions in even-$A$ 
Calcium isotopes are calculated in the full $fp$-model space using the nuclear 
shell model code NuShellX@MSU \cite{BROWN2014115, BROWN2001517}. The 
\textit{single} Gamow-Teller operator is applied two times sequentially on the 
ground state of the parent nucleus to obtain the DGT strength. This method is 
different from the method used in Refs.~\cite{MUTO199213, 
PhysRevLett.120.142502}. 

The properties of the DGT distribution are examined and limiting cases when the 
SU(4) holds or when the spin orbit-orbit coupling is put to zero are studied. 
DGT sum rules were derived in Refs.~\cite{VOGEL1988259, PhysRevC.40.936, 
MUTO199213}, and recently discussed in Ref.~\cite{PhysRevC.94.064325}. The DGT 
sum rules were used here as a tool to asses whether in our numerical 
calculations most of the DGT strength is found.

\section{Method of calculation}
\label{Method}
The notion of a DGT was introduced in Refs.~\cite{AUERBACH198977, ZHENG1990343}.
First of all, the nuclear shell model wave functions of the initial ground 
state, $J = 1^+$ intermediate states, and $J = 0^+, 2^+$ final states were 
obtained using the shell model code NuShellX@MSU \cite{BROWN2014115, 
BROWN2001517} with the FPD6 \cite{RICHTER1991325} and KB3G 
\cite{POVES2001157_KB3G} interactions, in the complete $fp$-model space. 
The maximum of the number of intermediate states is 1000. Table 
\ref{Nofinalstates} shows the total number of final states that is possible in 
Ti isotopes. If the total number of final states is larger than 5000, the 
calculations were done up to 5000 final states. As one will see later, that is 
enough to exhaust almost the total strength.
\begin{table}[b!]
\caption{The total number of final states in the $fp$-model space and $f$-model 
space including the $f_{7/2}$ and the $f_{5/2}$ orbits only. 
\label{Nofinalstates}}
{%
\newcommand{\mc}[3]{\multicolumn{#1}{#2}{#3}}
\begin{tabular}{|l|r|r|r|r|r|r|r|r|}
\hline
&\mc{2}{c|}{$^{42}$Ti} & \mc{2}{c|}{$^{44}$Ti} & 
\mc{2}{c|}{$^{46}$Ti} & \mc{2}{c|}{$^{48}$Ti} \\ 
\hline
$J^{\pi}_f$ & \mc{1}{c|}{$0^+$} & \mc{1}{c|}{$2^+$} & \mc{1}{c|}{$0^+$} & 
\mc{1}{c|}{$2^+$} & \mc{1}{c|}{$0^+$} & \mc{1}{c|}{$2^+$} & \mc{1}{c|}{$0^+$} & 
\mc{1}{c|}{$2^+$}\\ \hline
$fp$-space & 4 & 8 & 158 & 596 & 2343 & 9884 & 14177 & 61953\\ 
\hline
$f$-space & 2 & 1 & 29 & 99 & 180 & 741 & 446 & 1899\\ 
\hline
\end{tabular}
}%
\end{table}

\begin{figure}[h!]
\includegraphics[scale=0.8]{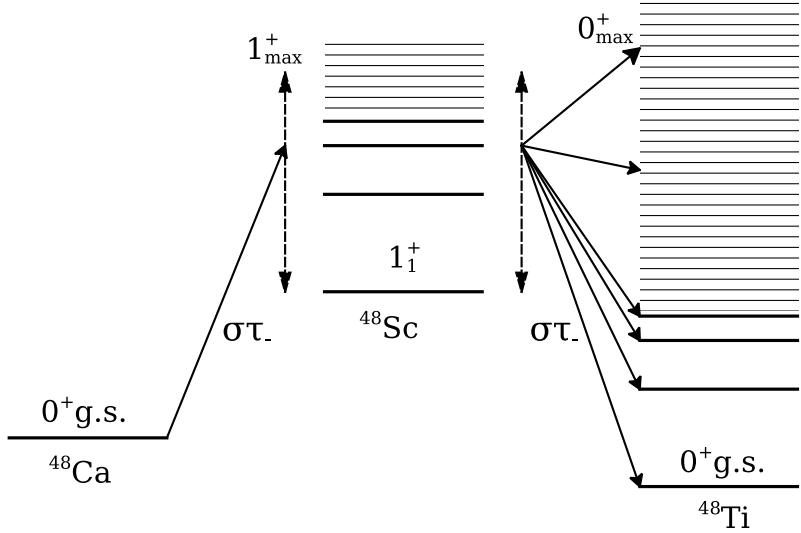}
\caption{Illustration of the method of calculation described in the text. The 
notations $1^+_{\text{max}}$ and $0^+_{\text{max}}$ are the highest states that 
can be reached in practice. \label{FigMethod}}
\end{figure}

After all wave functions were obtained, the single GT operator was applied two 
times sequentially. First, all transitions from the parent nucleus $0^+$ to all 
$1^+$ intermediate states are calculated and then all transitions from $1^+$ 
intermediate states to each $0^+$ or $2^+$ in the final nucleus are calculated. 
Fig.~\ref{FigMethod} illustrates the method of calculation.
The single GT operator is denoted as
\begin{equation}
 \bm{Y}_{\pm} = \sum_{i = 1}^A \bm{\sigma} t_\pm (i); \quad t_{\pm} = t_x \pm 
it_y,
\end{equation}
with $t_-n = p$ and $t_+ p = n$ where $2t_x$ and $2t_y$ are the Pauli isospin 
operators and $\bm{\sigma}$ is Pauli spin operator. Then the single GT 
transition amplitude from the initial state $|i\rangle$ to the final state 
$|f\rangle$ is
\begin{equation}
 M(GT_{\pm}; i \rightarrow f) = \frac{\langle f || \bm{Y}_{\pm} || i 
\rangle}{\sqrt{2J_i 
+ 1}},
\end{equation}
and the GT transition strength given by
\begin{equation}
 B(GT_{\pm}) = |M(GT_{\pm}; i \rightarrow f)|^2
\end{equation}
obeys the $"3(N-Z)"$ sum rule:
\begin{equation}
 \sum_f B(GT_{-}) - \sum_f B(GT_{+}) = S_{GT_-} - S_{GT_+} = 3(N - Z),
\end{equation}
where the $\sum_f$ means summing over all eigenstates of $J_fT_f$. 
Because the $fp$-model space is limited, only the valence neutrons 
participate in the calculation for Calcium isotopes. Therefore, we have 
$S_{GT_+} = 0$.

The dimensionless DGT transition amplitude is defined as
\begin{equation}
 M(DGT_{\pm}) = \sum_n M(GT_{\pm}; i\rightarrow n) M(GT_{\pm}; 
n\rightarrow f),
\end{equation}
where $n$ are the intermediate states. Note that this is a coherent sum. 
Finally, the DGT strength is given by
\begin{equation}
 B(DGT_{\pm}) = |M(DGT_{\pm})|^2,
\end{equation}
or in more detail
\begin{equation}
  B(\text{DGT}^-; i \rightarrow n \rightarrow  f) = \frac{1}{2J_i + 1} 
\left| \sum_n 
\left\langle f \left| \left| \sum_i \bm{\sigma}_i t_{-}(i) \right| \right| n 
\right\rangle  \left\langle n \left| \left| \sum_j \bm{\sigma}_j t_{-}(j) 
\right| \right| i \right\rangle \right|^2.
\end{equation}
Note that $B(DGT^-; i \rightarrow n \rightarrow f)$ depends on $J_f$ with $J_f 
= 0$ and 2 only. The matrix element in the case $J_f = 1$ 
vanishes because the DGT operator changes sign under the interchange of 
coordinates of two particles.

The DGT sum rule for $J_f = 0$ is given in Refs.~\cite{VOGEL1988259, 
PhysRevC.40.936} and $J_f = 0, 2$ are given in Refs.~\cite{MUTO199213, 
PhysRevC.94.064325}. In summary, the generalization of the sum rules for DGT 
operators are:
\begin{eqnarray} \label{DGTsumrule}
 S_{\text{DGT}}^{J_f = 0} =  6(N-Z)(N-Z+1) - 2\Delta, \nonumber \\
 S_{\text{DGT}}^{J_f = 2} = 30(N-Z)(N-Z-2) + 5\Delta,
\end{eqnarray}
where
$ \Delta = \sqrt{2} \langle 0| [\bm{Y}_+ \times \bm{Y}_-]^{(1)} \cdot 
\bm{\varSigma} - \bm{\varSigma} \cdot [\bm{Y}_- \times \bm{Y}_+]^{(1)} | 
0\rangle,$ with $\bm{\varSigma} = \sum_i \bm \sigma (i)$. There is a factor 
of three difference between our work and the work in Refs.~\cite{MUTO199213, 
PhysRevLett.120.142502} because the spin operator is not projected in our 
calculation. The first terms of the sum rules depend only on $N$ and $Z$, and 
the second terms ($2\Delta$ or $5\Delta$) need to be calculated separately. The 
sign of the second term makes the first one to be the upper limit for $J_f = 
0^+$ and lower limit for $J_f = 2^+$.

\section{Results and discussions}
\begin{figure}
\includegraphics[scale=0.7]{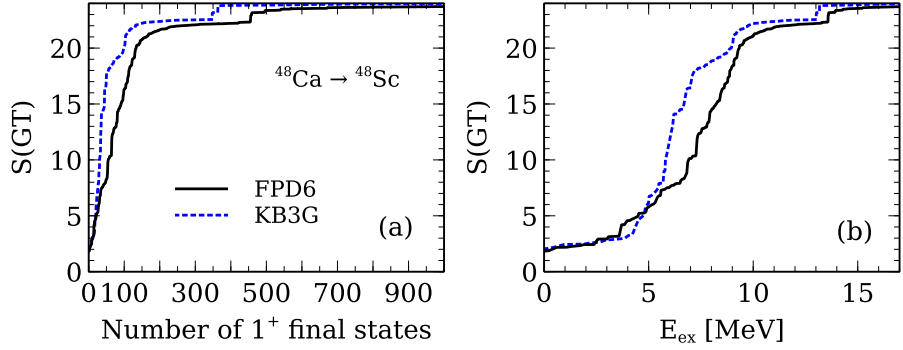}
\caption{The cumulative sum of the single Gamow-Teller strength B(GT) as 
a function of the number of $1^+$ states (a) and excitation energies (b) of the 
intermediate nucleus $^{48}$Sc. The calculation used FPD6 and KB3G 
interactions in the $fp$-model space. \label{SGTCa48}}
\end{figure}

It is well-known that the \textit{single} GT strength is quenched (see 
Ref.~\cite{RevModPhys.64.491}). In the shell model calculations, GT 
strength is fragmented. This is demonstrated in the case of $^{48}$Ca in 
Fig.~\ref{SGTCa48} as an example. The results were obtained with two different 
interactions that are often used in the $fp$-model space FPD6 and KB3G 
interactions. Within the range of about 17 MeV excitation energy, $S_{GT_-}$ is 
approximately 24 exhausting the $``3(N-Z)"$ sum rule ($S_{GT_+} = 0$ in our 
calculation). The cumulative sum of the single GT strength S(GT) as a function 
of number of $^{48}$Sc states is shown in Fig.~\ref{SGTCa48}. One can expect 
that there are about 500 intermediate $1^+$ states in $^{48}$Sc that actually 
contribute to the final results of the DGT strength although the total number of 
$1^+$ states in this nucleus is many thousands.

\begin{table}[b!]
\caption{The properties of DGT transition using FPD6 and KB3G interactions in 
the complete $fp-$model space. $S(\text{DGT})$ is the DGT total strength with 
the FPD6 interaction. $B_1$ is the transition from the g.s. of the parent 
nucleus to the first $J^+$ state of the final nucleus. $\overline{E}$ (MeV) is 
the average energy (Eq.~(\ref{Eaverage})). \label{SDGT}}
{%
\newcommand{\mc}[3]{\multicolumn{#1}{#2}{#3}}
\begin{center}
\begin{tabular}{|l|r|r|r|r|r|r|r|}
\hline
 Initial nucleus &\mc{2}{c|}{$^{42}$Ca} & \mc{2}{c|}{$^{44}$Ca} & 
\mc{2}{c|}{$^{46}$Ca} & $^{48}$Ca \\ 
\hline
$J^{\pi}_f$ & \mc{1}{c|}{$0^+$} & \mc{1}{c|}{$2^+$} & \mc{1}{c|}{$0^+$} & 
\mc{1}{c|}{$2^+$} & 
\mc{1}{c|}{$0^+$} & \mc{1}{c|}{$2^+$} & \mc{1}{c|}{$0^+$} \\ 
\hline
$S(\text{DGT})$ & \phantom{1} 28.1 & \phantom{1} 19.5 & 102.0 & 284.0 & 223.7 & 
752.6 & 385.0 \\
Sum rule & $\leq 36$ & $\geq 0$ & $\leq 120$ & $\geq 240$ & 
$\leq 252$ & $\geq 720$ & $\leq 432$ \\ 
$B_1$ (FPD6)
& 16.172 & 6.117 & 0.654 & 0.000 & 0.201 & 0.017 & 0.109 \\
$B_1$ (KB3G)
& 17.010 & 5.942 & 0.895 & 0.119 & 0.182 & 0.057 & 0.072 \\
$\overline{E} $ (FPD6)
& 6.1 & 4.8 & 16.3 & 13.2 & 21.2 & 18.0 & 24.6 \\
$\overline{E} $ (KB3G)
& 6.1 & 5.5 & 14.7 & 12.2 & 19.0 & 16.9 & 21.9 \\ 
\hline
\end{tabular}
\end{center}
}%
\end{table}
For the study of the DGT, first, we calculated the sum rule using it as a tool 
to asses whether in our numerical calculations most of the DGT strength is 
found. The $\Delta$ in Eq.~(\ref{DGTsumrule}) can be obtained 
directly by subtracting from total sum the first term that depends only on $N$ 
and $Z$ (see Table \ref{SDGT}).
In Ref.~\cite{PhysRevC.40.936}, $\Delta$ was related to the magnetic dipole 
transition $S(M1)$. Table I in Ref.~\cite{PhysRevC.40.936} and Table I in 
Ref.~\cite{PhysRevC.94.064325} gave the values of the sum rules for even-$A$ 
isotopes including Calcium isotopes. Our results given in Table \ref{SDGT} are 
in agreement with them (Note that there is a factor of three difference between 
our work and Ref.~\cite{PhysRevC.94.064325}). It means we exhaust all the 
DGT strength in the study. Obviously, the total DGT strength does not depend 
on the choice of interaction.

\begin{figure}
\includegraphics[scale=0.7]{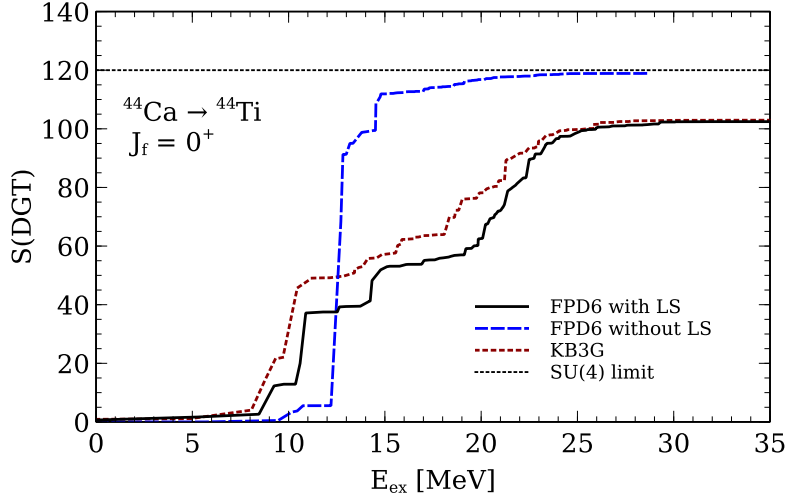}
\caption{The cumulative sum of the double Gamow-Teller strength $B(\rm{DGT}; 
0^+ \rightarrow 0^+)$ in $^{44}$Ca.\label{DGTCa44cumsum0}}
\end{figure}
\begin{figure}
\includegraphics[scale=0.7]{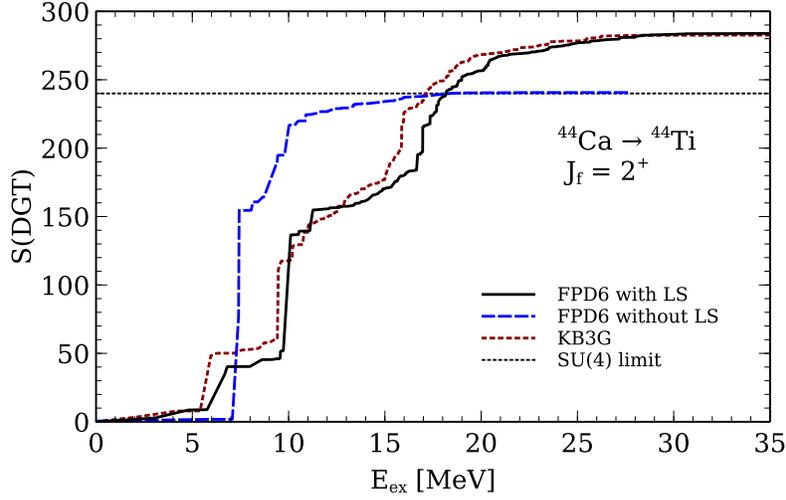}
\caption{The cumulative sum of the double Gamow-Teller strength $B(\rm{DGT}; 
0^+ \rightarrow 2^+)$ in $^{44}$Ca.\label{DGTCa44cumsum2}}
\end{figure}
\begin{figure}
\includegraphics[scale=0.7]{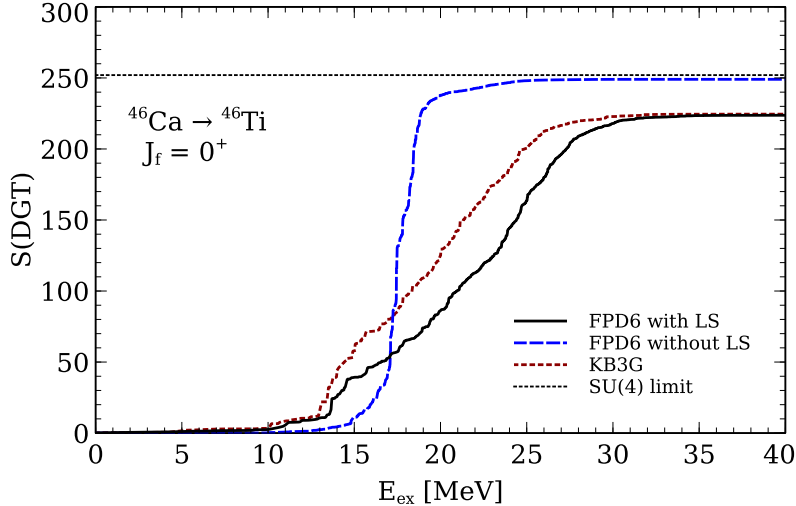}
\caption{The cumulative sum of the double Gamow-Teller strength $B(\rm{DGT}; 
0^+ \rightarrow 0^+)$ in $^{46}$Ca.\label{DGTCa46cumsum0}}
\end{figure}
\begin{figure}
\includegraphics[scale=0.7]{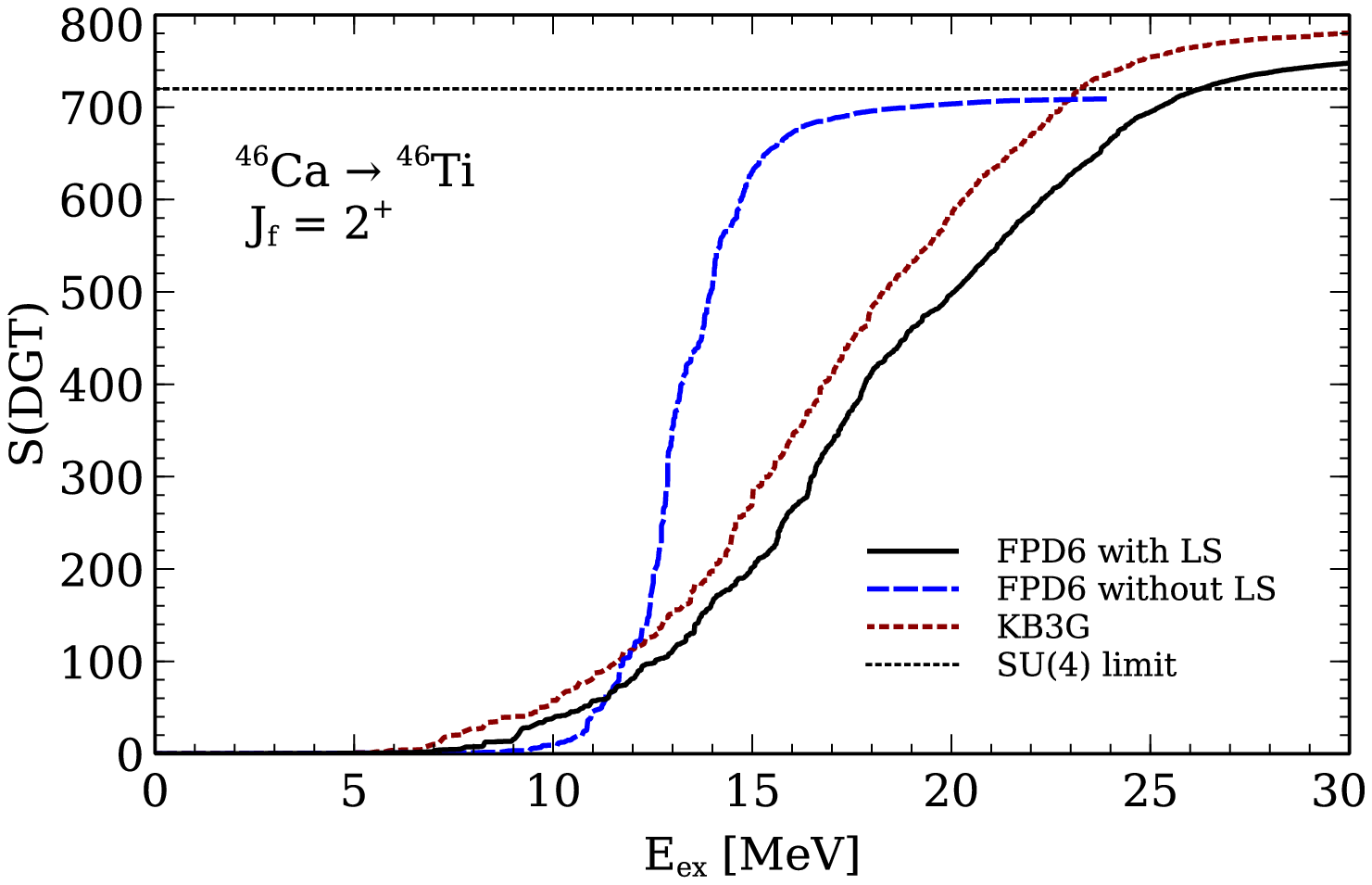}
\caption{The cumulative sum of the double Gamow-Teller strength $B(\rm{DGT}; 
0^+ \rightarrow 2^+)$ in $^{46}$Ca.\label{DGTCa46cumsum2}}
\end{figure}
\begin{figure}[b!]
\includegraphics[scale=0.7]{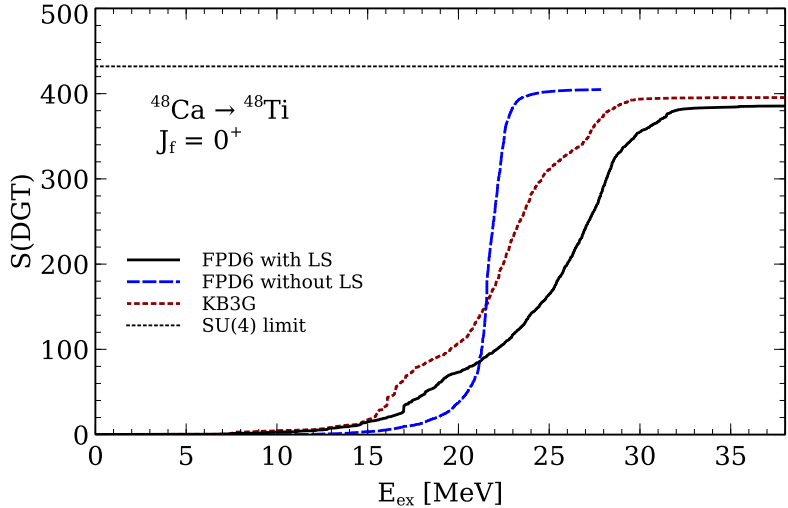}
\caption{The cumulative sum of the double Gamow-Teller strength $B(\rm{DGT}; 
0^+ \rightarrow 0^+)$ in $^{48}$Ca.\label{DGTCa48cumsum0}}
\end{figure}

Because all the strengths are obtained, we can show not only the values of the 
total sum but also the cumulative sums of the DGT. The cumulative sums are 
given in Fig.~\ref{DGTCa44cumsum0}, and Fig.~\ref{DGTCa44cumsum2} for 
$^{44}$Ca, Fig.~\ref{DGTCa46cumsum0}, and Fig.~\ref{DGTCa46cumsum2} for 
$^{46}$Ca, and Fig.~\ref{DGTCa48cumsum0} for $^{48}$Ca. In these figures, the 
solid lines are the shell model calculations described in Section \ref{Method} 
using the FPD6 interaction. They are denoted as ``FPD6 with LS''. The results 
using the KB3G interaction are also shown as dotted lines. The long-dash lines 
are the calculation with the FPD6 interaction in the SU(4) limit. The SU(4) 
limit in our work is restored approximately by making the $f_{5/2}$ and 
$f_{7/2}$; $p_{1/2}$ and $p_{3/2}$ degenerate following 
Ref.~\cite{PhysRevC.96.044319}.
It means there is no spin-orbit coupling and therefore they were denoted as 
``FPD6 without LS''. We want to show that in the SU(4) limit, the cumulative 
sums approach the horizontal lines (denoted as the ``SU(4) limit'') that 
represent the values of the terms that depend only on $N$ and $Z$ in 
Eq.~(\ref{DGTsumrule}). It is in agreement with the fact that $\Delta$ vanishes 
in the SU(4) limit according to Ref.~\cite{PhysRevC.39.2370}. 
In the cases of $^{44}$Ca (Fig.~\ref{DGTCa44cumsum0}, and 
Fig.~\ref{DGTCa44cumsum2}), the sum rules are totally exhausted because all 
intermediate states and all final states can be taken into account. In the 
cases of $^{46}$Ca, and $^{48}$Ca 
(Figs.~\ref{DGTCa46cumsum0}--\ref{DGTCa48cumsum0}), the cumulative 
sums are still increasing. For the case of the DGT to the $2^+$ in $^{48}$Ca, 
we choose not to do the calculation because the total number of final states 
are too large. The result is not convergent using the standard NUSHELLX@MSU 
code \cite{BrownCommu}.

Most of the sum rule is satisfied, and therefore the entire distributions of 
DGT strength now can be discussed. We remind that Ref.~\cite{MUTO199213} showed 
the entire DGT distributions for even-$A$ Neon and recently 
Ref.~\cite{PhysRevLett.120.142502} showed the result for $^{48}$Ca for the 
first time. For the lightest nucleus, $^{42}$Ca, the DGT distributions with FPD6 
and KB3G interactions are shown in Table \ref{42J0} and \ref{42J2}. The 
difference between the results of the two interactions is not large.
\begin{table}[b!]
\caption{$B(\rm{DGT}; 0^+ \rightarrow 0^+)$ for $^{42}$Ca using FPD6 and KB3G 
interactions in the complete $fp$-model space. $E_{ex}$ (MeV) is the excited 
energy of $^{42}$Ti. \label{42J0}}
 \begin{center}
\begin{tabular}[t]{|rrr||rrr|}
\hline
 & FPD6 & &  & KB3G &\\
\hline
$E_{ex}$&	$B(\rm{DGT})$	&	$S(\rm{DGT})$	& $E_{ex}$&	
$B(\rm{DGT})$	&	$S(\rm{DGT})$	\\
\hline
0.0	&	16.172	&	16.172	&	0.0	&	17.010	&	
17.010	\\
6.0	&	0.442	&	16.614	&	5.7	&	0.281	&	
17.291	\\
10.9	&	0.782	&	17.396	&	11.3	&	0.120	&	
17.411	\\
14.9	&	10.692	&	28.088	&	15.4	&	11.085	&	
28.496	\\
\hline
\end{tabular}
\end{center}
\end{table}
\begin{table}[t!]
\caption{The same as Table \ref{42J0}, but for $B(\rm{DGT}; 0^+ \rightarrow 
2^+)$ \label{42J2}}
 \begin{center}
\begin{tabular}[t]{|rrr||rrr|}
\hline
 & FPD6 & &  & KB3G &\\
\hline
$E_{ex}$&	$B(\rm{DGT})$	&	$S(\rm{DGT})$	& $E_{ex}$&	
$B(\rm{DGT})$	&	$S(\rm{DGT})$	\\
\hline
0.0	&	6.117	&	6.117	&	0.0	&	5.942	&	
5.943	\\
2.3	&	1.536	&	7.653	&	2.6	&	0.520	&	
6.463	\\
5.1	&	0.125	&	7.778	&	5.2	&	0.009	&	
6.472	\\
6.6	&	5.523	&	13.301	&	7.2	&	1.355	&	
7.827	\\
7.3	&	4.916	&	18.217	&	7.8	&	9.679	&	
17.506	\\
9.7	&	0.071	&	18.288	&	10.1	&	0.188	&	
17.694	\\
11.6	&	0.039	&	18.327	&	11.9	&	0.017	&	
17.711	\\
14.2	&	1.148	&	19.475	&	14.2	&	1.047	&	
18.757	\\
\hline
\end{tabular}
\end{center}
\end{table}
In the case $B(\text{DGT}; 0^+ \rightarrow 0^+)$ of $^{42}$Ca, the transition 
from the g.s. of the parent nucleus to the first $J^+$ state of the final 
nucleus ($B_1$) is large because the g.s. of $^{42}$Ti is the DIAS of the g.s. 
of $^{42}$Ca. Moreover, in the SU(4) limit the g.s. of $^{42}$Ti absorbs all 
the DGT strength (36) following Refs.~\cite{VOGEL1988259, PhysRevC.40.936}.

\begin{figure}
\includegraphics[scale=0.7]{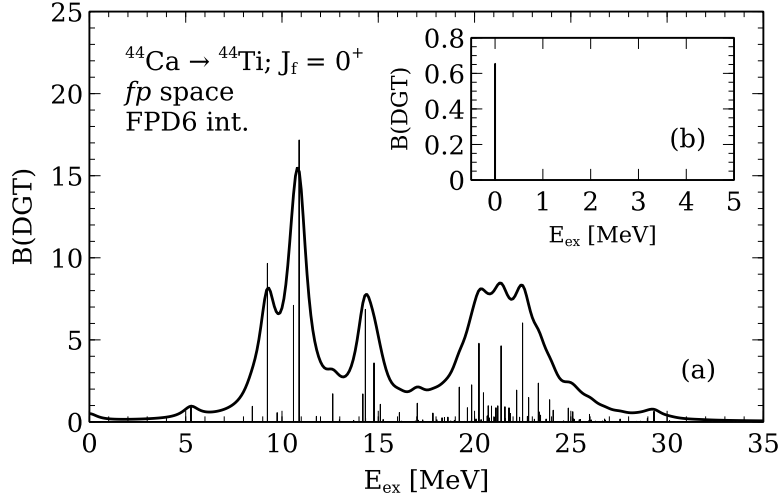}
\caption{$B(\rm{DGT}; 0^+ \rightarrow 0^+)$ for $^{44}$Ca as the function of 
the excitation energy of the final nucleus $^{44}$Ti (a). The DGT 
transitions to low-lying states are also shown (b). The strengths 
were smoothed by using Lorentzian averaging with the width of 1 MeV. 
\label{DGTCa441}}
\end{figure}

\begin{figure}
\includegraphics[scale=0.7]{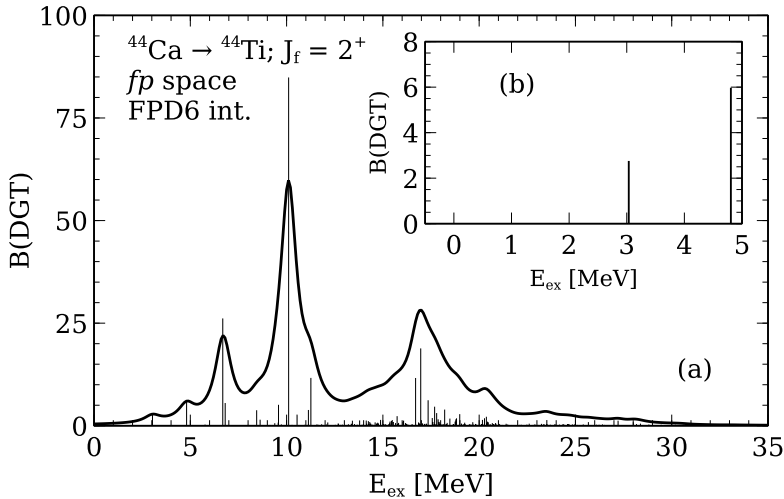}
\caption{The same as Fig.~\ref{DGTCa441}, but for $B(\rm{DGT}; 0^+ \rightarrow 
2^+)$. \label{DGTCa442}}
\end{figure}

\begin{figure}
\includegraphics[scale=0.7]{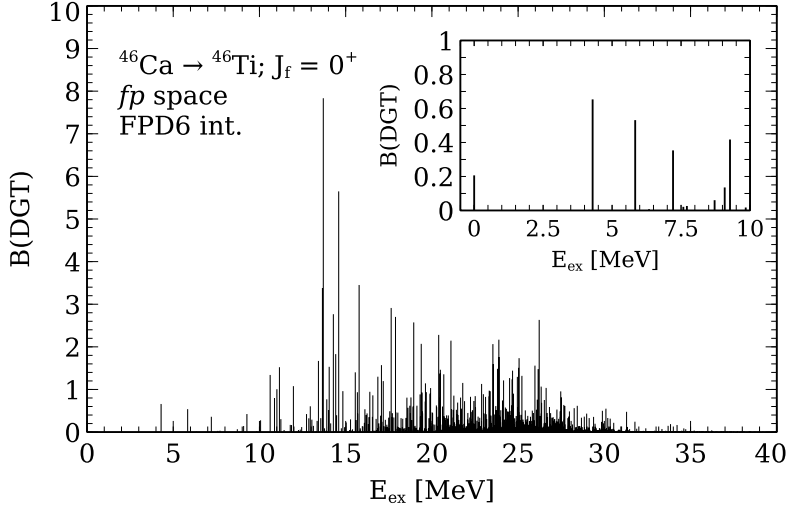}
\caption{$B(\rm{DGT}; 0^+ \rightarrow 0^+)$ in $^{46}$Ca.\label{DGTCa461}}
\end{figure}

\begin{figure}
\includegraphics[scale=0.7]{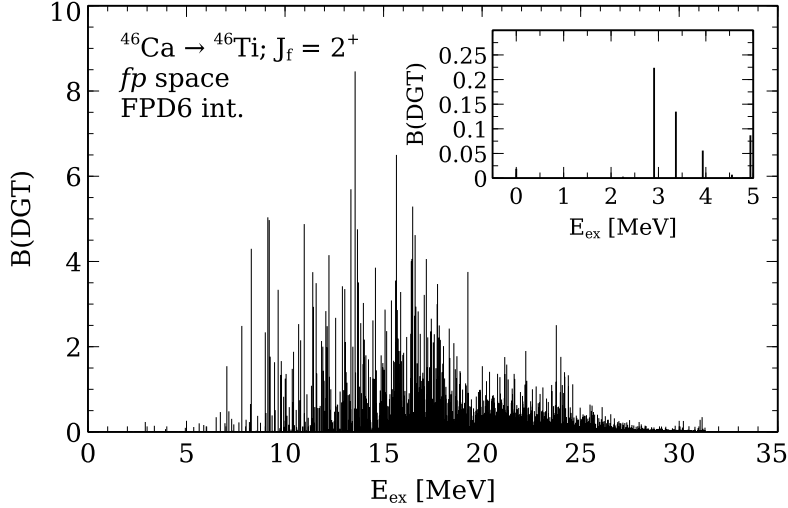}
\caption{$B(\rm{DGT}; 0^+ \rightarrow 2^+)$ in $^{46}$Ca.\label{DGTCa462}}
\end{figure}

\begin{figure}
\includegraphics[scale=0.7]{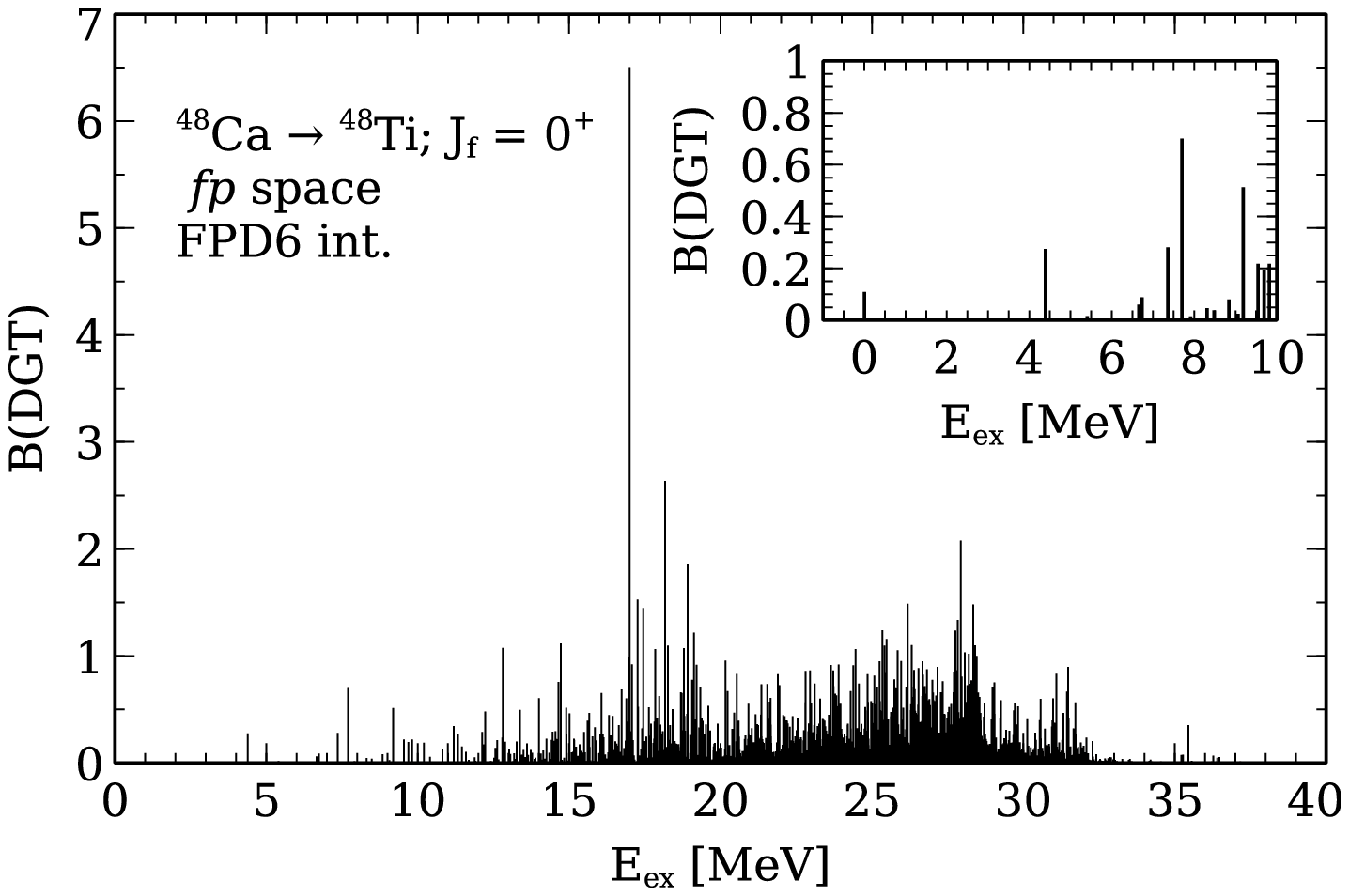}
\caption{$B(\rm{DGT}; 0^+ \rightarrow 0^+)$ in $^{48}$Ca.\label{DGTCa48J0}}
\end{figure}

The DGT distributions are drawn in Figs.~\ref{DGTCa441}--\ref{DGTCa48J0}. 
They contain inserts which show the DGT strength in the low-lying states of 
$^{44,46,48}$Ti. $B_1$ is a very tiny fraction of the total strength. For 
example, the strength in the ground state of $^{48}$Ti is only $3\times10^{-4}$ 
of the total strength (see Table \ref{SDGT}). This strength enters in the 
calculation of the $\beta\beta$ decay. In Ref.~\cite{PhysRevLett.120.142502}, 
it is pointed out that a very good linear correlation between the DGT 
transition to the ground state of the final nucleus and the $0\mu\beta\beta$ 
decay matrix element exists.

When the strengths are spread by using the same Lorentzian averaging 
with the width of 1 MeV to simulate the experimental energy resolution, the 
results show that the DGT distributions are not single-peaked. The 
distributions have at least two peaks and in some nuclei as many as four major 
peaks. We should remind that the single GT resonances have at least two peaks 
\cite{GOODMAN1982241}.

\begin{figure}
\includegraphics[scale=0.7]{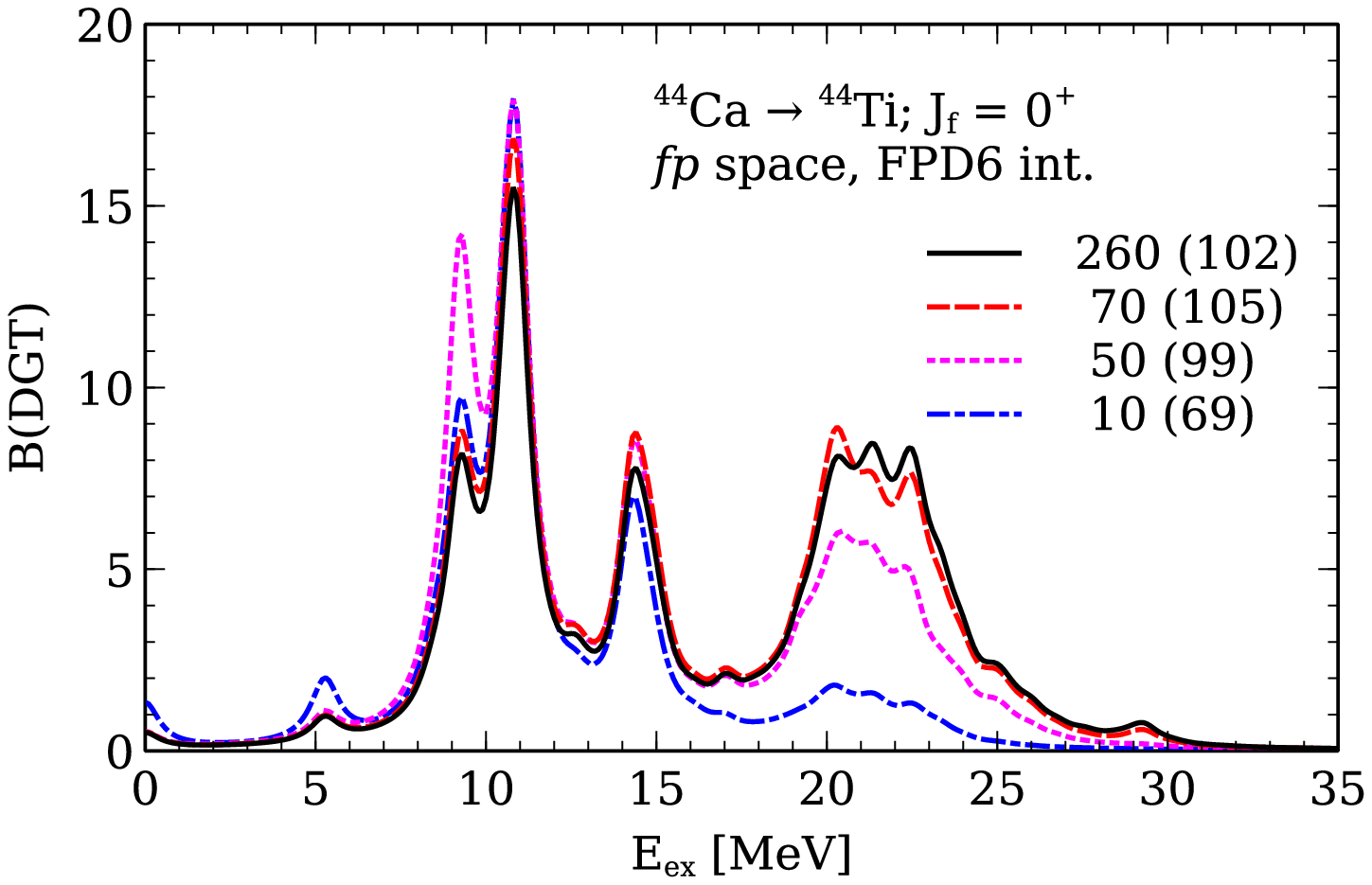}
\caption{The dependence on the number of intermediate states of 
$B(\rm{DGT}; 0^+ \rightarrow 0^+)$ in $^{44}$Ca. The numbers in 
parentheses are the corresponding total strengths. \label{DGTCa44cut0}}
\end{figure}

\begin{figure}
\includegraphics[scale=0.7]{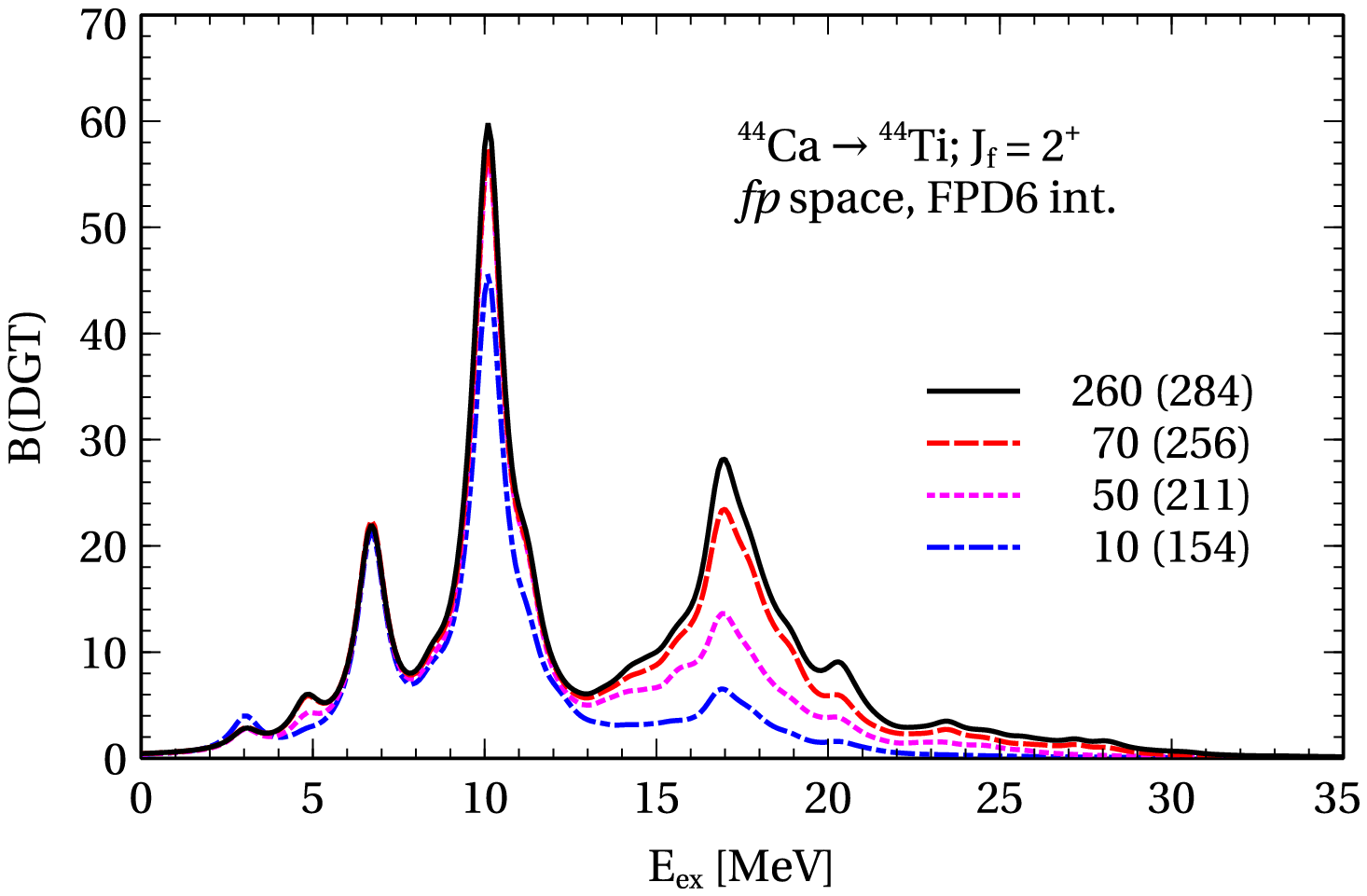}
\caption{The same as Fig.~\ref{DGTCa44cut0}, but for 
$B(\rm{DGT}; 0^+ \rightarrow 2^+)$ in $^{44}$Ca. \label{DGTCa44cut2}}
\end{figure}

\begin{figure}
\includegraphics[scale=0.7]{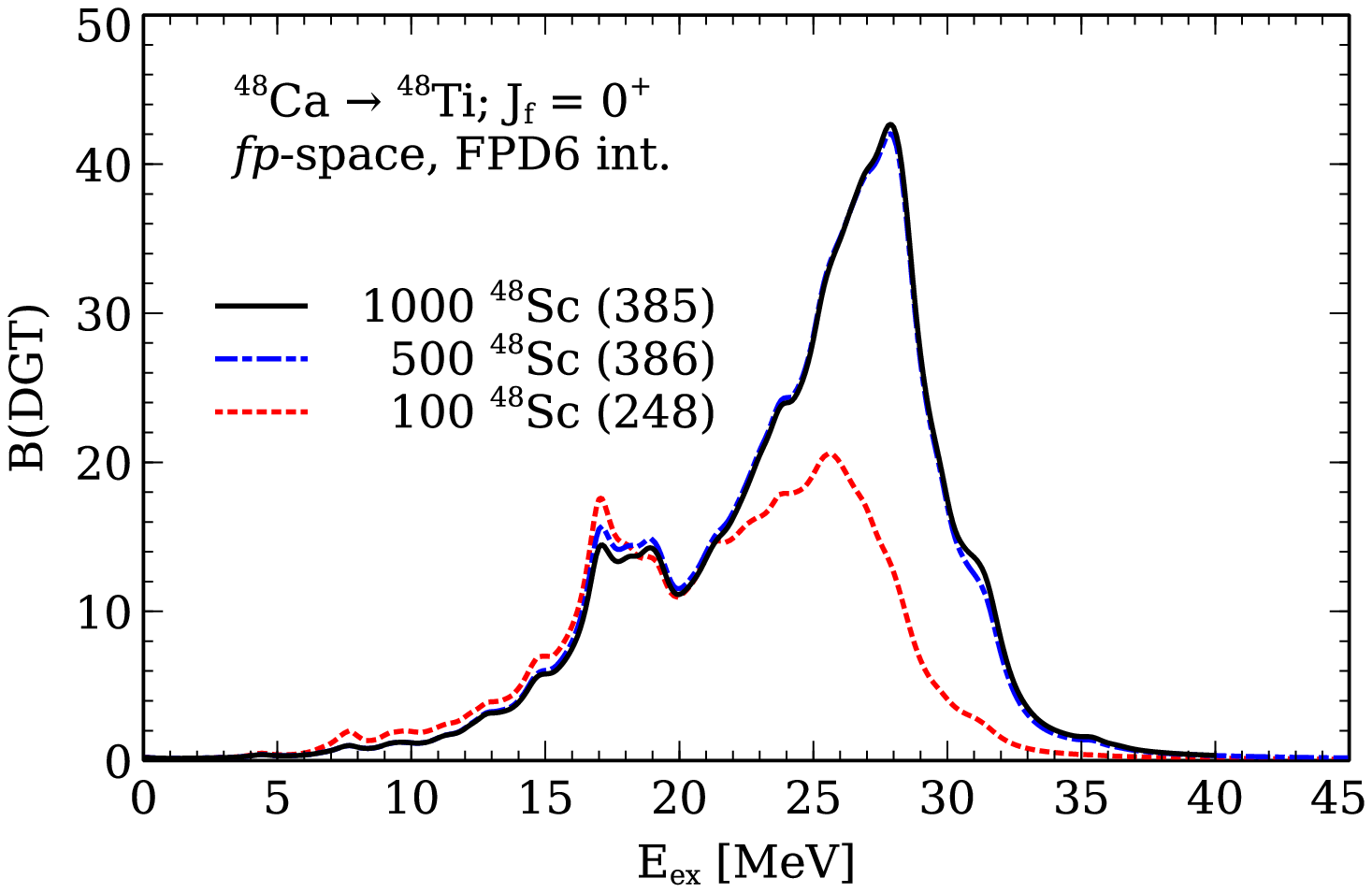}
\caption{The same as Fig.~\ref{DGTCa44cut0}, but for 
$B(\rm{DGT}; 0^+ \rightarrow 0^+)$ in $^{48}$Ca.\label{DGTCa48cut0}}
\end{figure}

\begin{figure}
\includegraphics[scale=0.7]{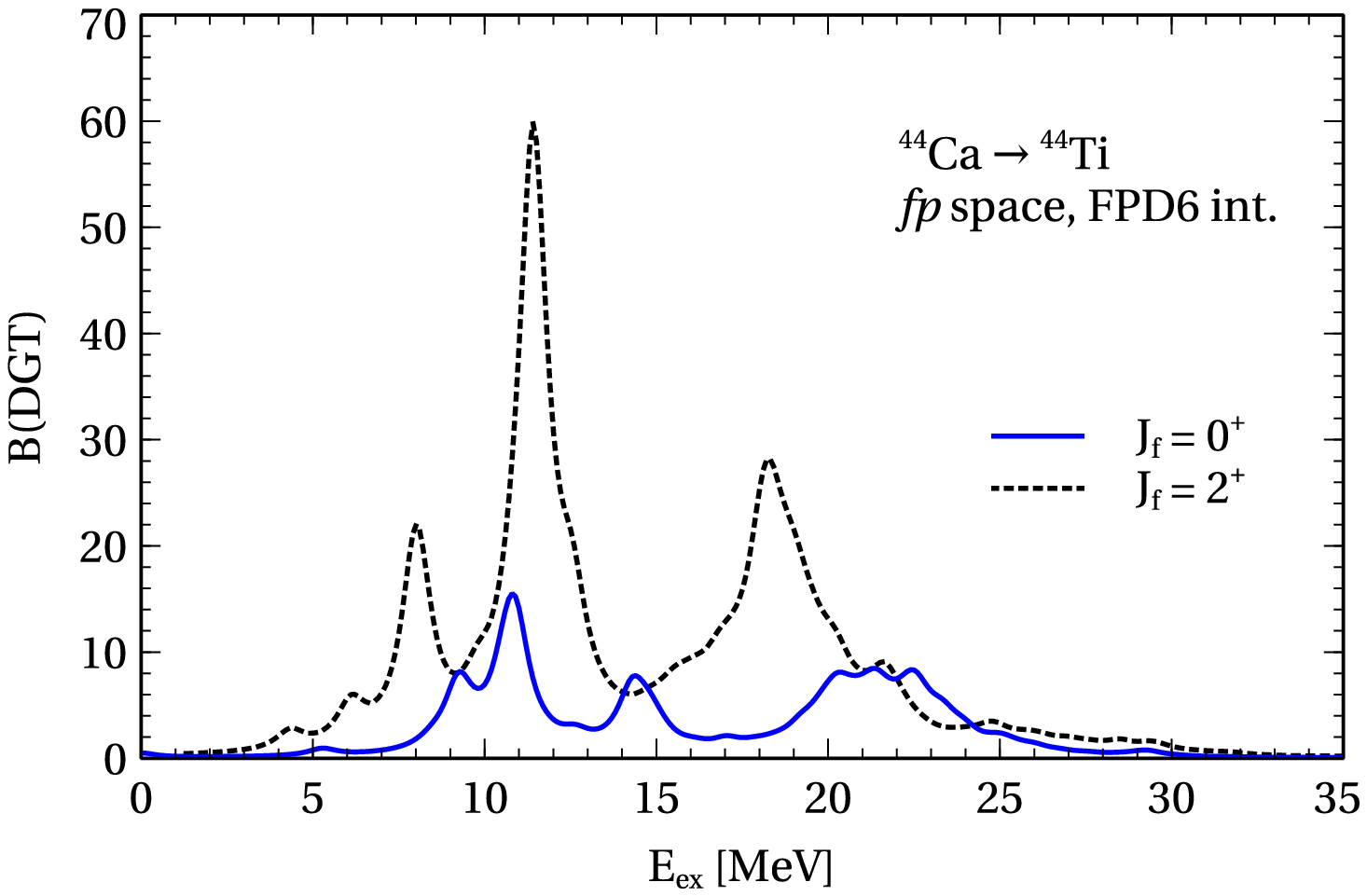}
\caption{$B(\rm{DGT}; 0^+ \rightarrow 0^+)$ and $B(DGT; 0^+ \rightarrow 2^+)$ 
in $^{44}$Ca.\label{DGTCa445}}
\end{figure}

\begin{figure}
\includegraphics[scale=0.7]{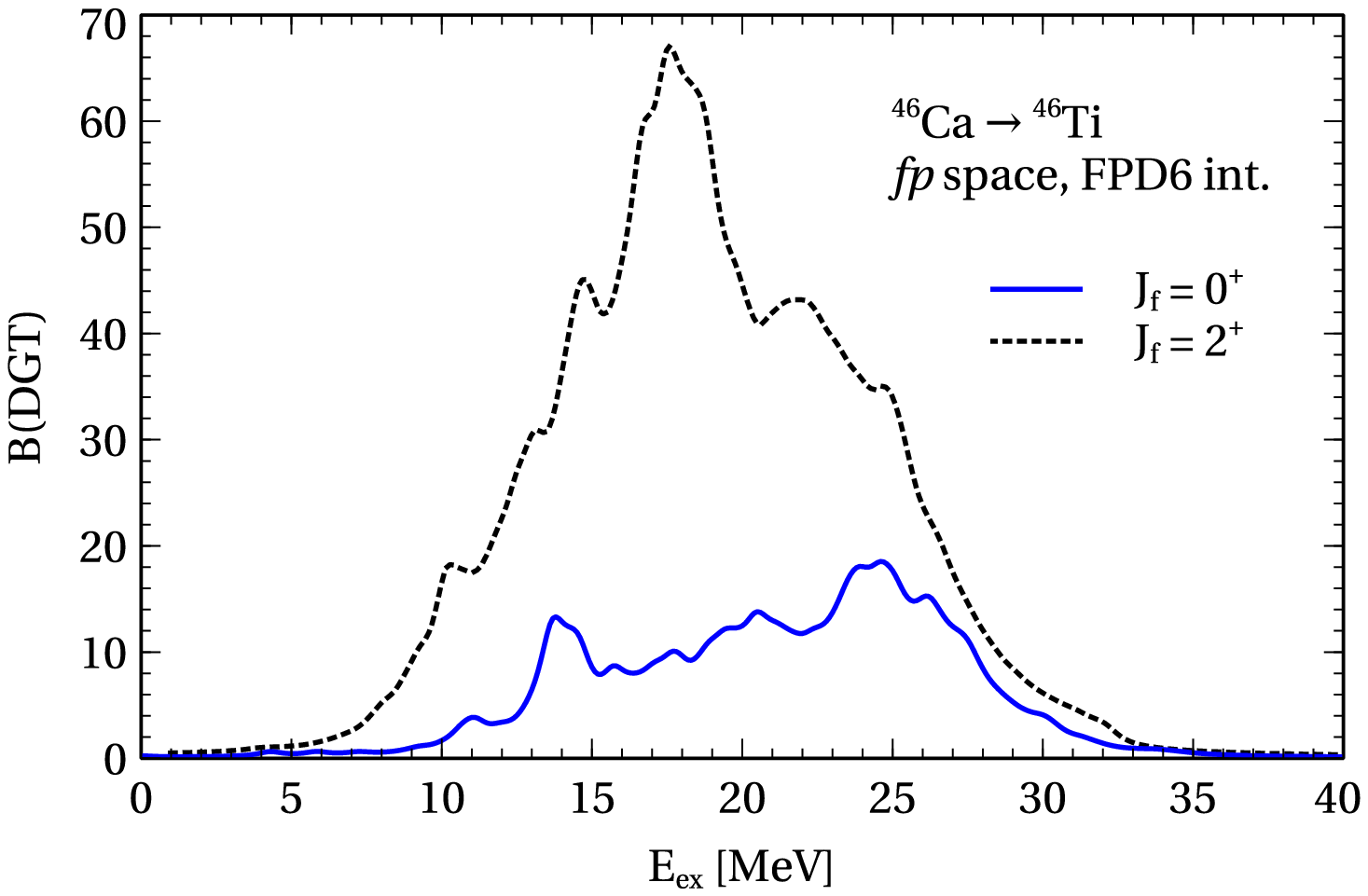}
\caption{The same as Fig.~\ref{DGTCa445} but for $^{46}$Ca. \label{DGTCa467}}
\end{figure}

Figs.~\ref{DGTCa44cut0}--\ref{DGTCa48cut0} show the dependence of the DGT
distributions on the number of intermediate states. We can see that about 100 
intermediate states actually contribute to final results in the cases of 
$^{44}$Ca (Figs.~\ref{DGTCa44cut0}--\ref{DGTCa44cut2}). Although the total 
number of intermediate states in heavier isotopes, including $^{48}$Ca is 
many thousands, about 500 intermediate states actually contribute to final 
results (Fig.~\ref{DGTCa48cut0}). The sum rule is useful to determine this 
number (We remind that the number of intermediate states involved in the 
calculations for $0\nu\beta\beta$ decay is smaller (see 
Ref.~\cite{PhysRevC.88.064312})).
Fig.~\ref{DGTCa445} and Fig.~\ref{DGTCa467} show the DGT transitions to 
$J_f^\pi = 0^+$ together with the transition to $J_f^\pi = 2^+$ in $^{44}$Ca 
and $^{46}$Ca. As one can see the DGT transitions to $J_f^\pi = 2^+$ are 
larger than the transitions to $J_f^\pi = 0^+$.

\begin{figure}
\includegraphics[scale=0.7]{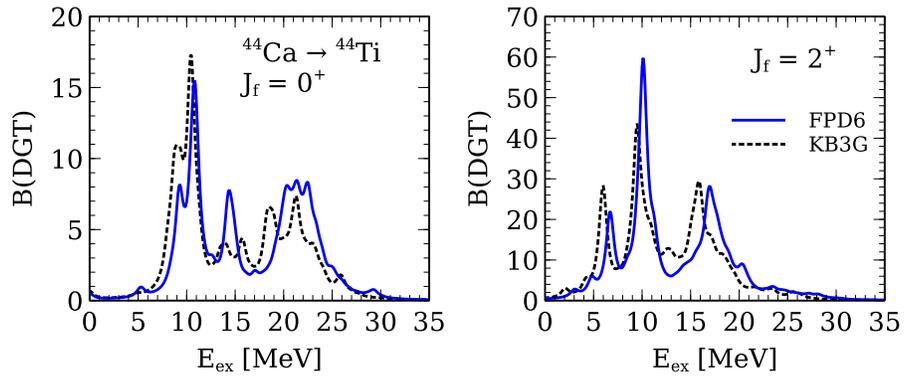}
\caption{$B(\rm{DGT}; 0^+ \rightarrow 0^+; 2^+)$ in $^{44}$Ca using FPD6 and 
KB3G interactions.\label{DGTCa442int2J}}
\end{figure}

\begin{figure}[b!]
\includegraphics[scale=0.7]{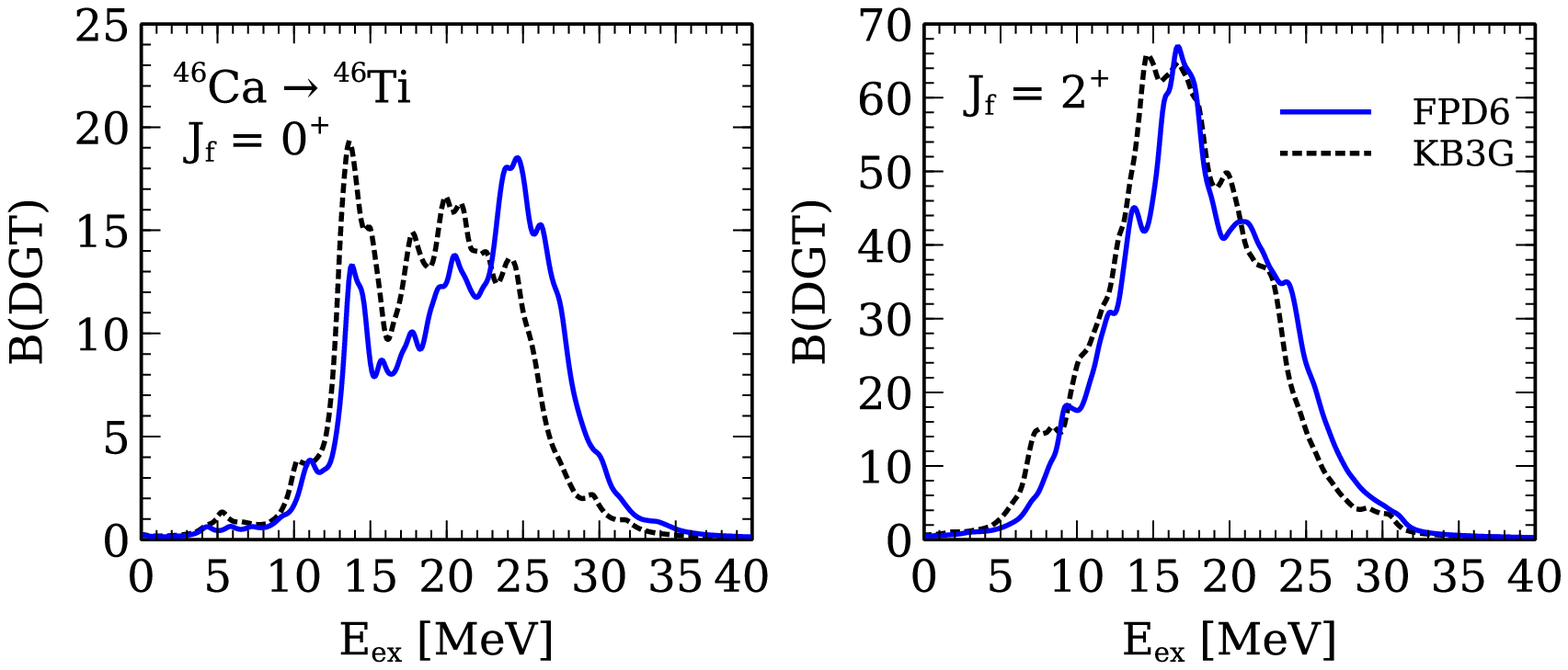}
\caption{The same as Fig.~\ref{DGTCa442int2J}, but for 
$^{46}$Ca.\label{DGTCa462int2J}}
\end{figure}

\begin{figure}[b!]
\includegraphics[scale=0.7]{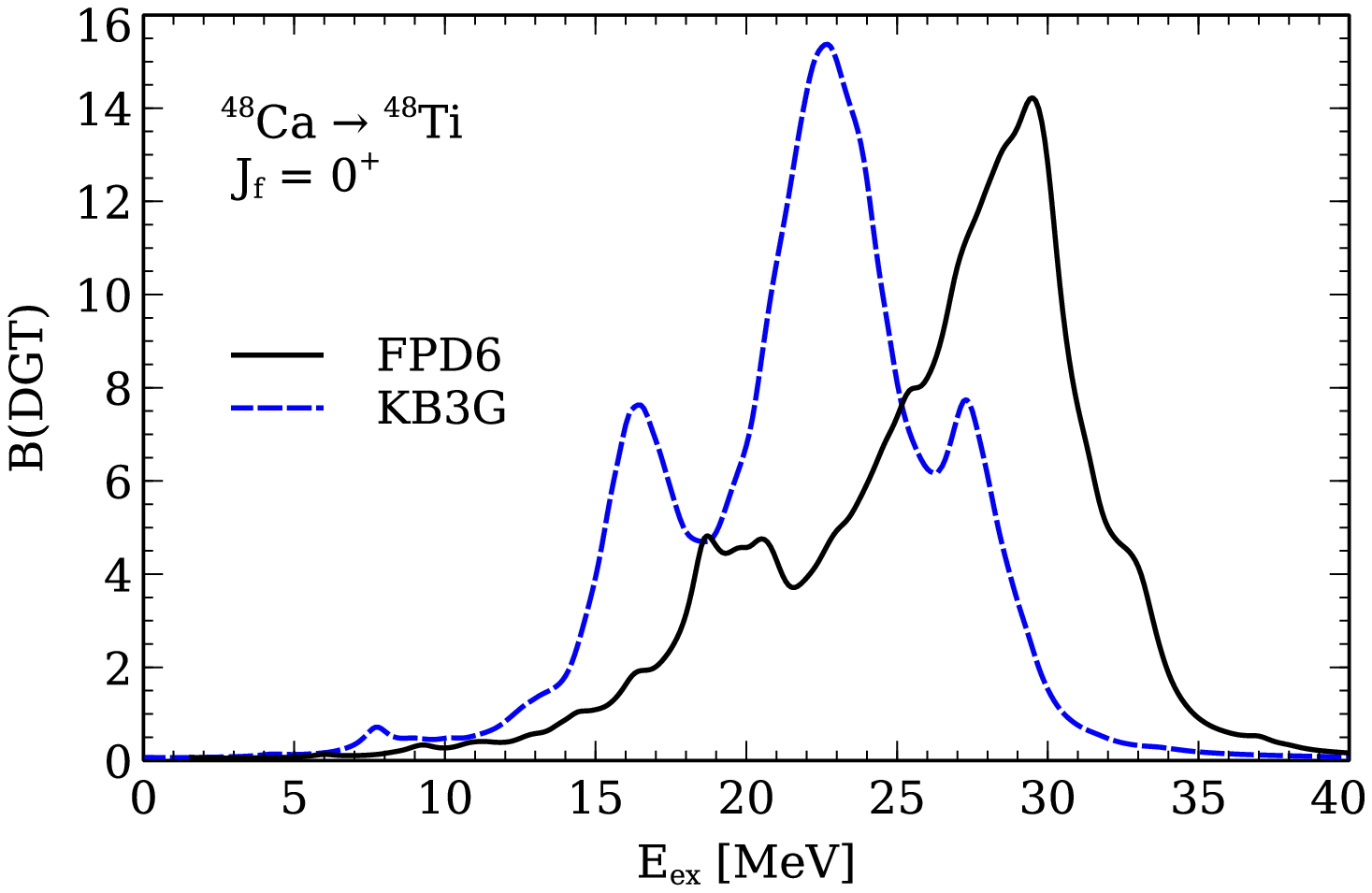}
\caption{$B(\rm{DGT}; 0^+ \rightarrow 0^+)$ in $^{48}$Ca using FPD6 and KB3G 
interactions. The strengths are spread with the width 1 MeV and divided by the 
factor of 3 for comparing to Ref~.\cite{PhysRevLett.120.142502}. 
\label{DGTCa482int}}
\end{figure}

The centroid or average energies of the DGT distributions defined by
\begin{equation}\label{Eaverage}
 \overline{E} = \frac{\sum_f E_f B_f(DGT_{-})}{\sum_f B_f(DGT_{-})}
\end{equation} 
are given in Table \ref{SDGT}. 
In $^{46}$Ti, with the FPD6 interaction for example, the average energy for the 
$J = 0^+$ is $\overline{E} = 21.2$ MeV and for the $J = 2^+$ it is lower, 
$\overline{E} = 18.0$ MeV. In $^{48}$Ti the average energy $J = 0^+$ is 
$\overline{E} = 24.6$ MeV.
In Ref.~\cite{PhysRevLett.120.142502}, a simple relation between the average 
energy of the $^{48}$Ca DGT giant resonance and $0\mu\beta\beta$ decay nuclear 
matrix element was pointed out. The authors conclude that the uncertainties due 
to the nuclear interaction in the calculation of the DGT distribution in 
$^{48}$Ca are relatively under control.
In our work, Figs.~\ref{DGTCa442int2J}--\ref{DGTCa482int} show the DGT 
distributions are calculated with FPD6 and KB3G interactions. We see that the 
distributions and the average energies (see Table \ref{SDGT}) using FPD6 and 
KB3G are similar. Our calculated distribution for $^{48}$Ca is in agreement with 
the recent result using the same KB3G interaction but a different method (when 
the factor of three is taken into account). As one can see in 
Fig.~\ref{DGTCa482int} the DGT giant resonance in $^{48}$Ca is at the energy 
around 20-30 MeV.
In a recent paper \cite{Ejiri2017} the experimental results for the 
double charge-exchange reaction $^{56}$Fe($^{11}$B, $^{11}$Li) were presented. 
In this reaction, several resonances were excited in agreement with the pion 
DCX studies. In addition, there is a peak at 25 MeV excitation, that the authors 
indicate that it could be the DGT resonance.

\begin{figure}
\includegraphics[scale=0.9]{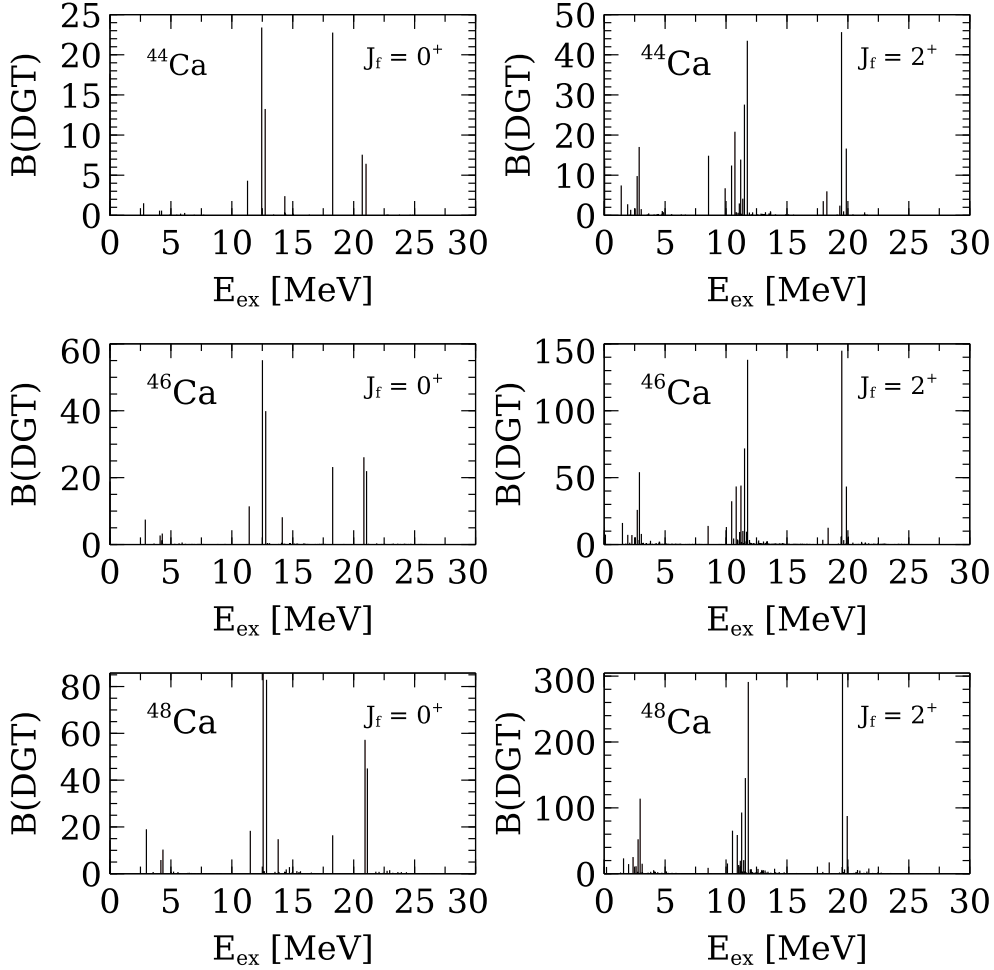}
\caption{The DGT distributions in the $f$-model space (including the $f_{7/2}$ 
and the $f_{5/2}$ orbits only) in Calcium isotopes using the FPD6 interaction. 
\label{DGTfspace}}
\end{figure}

In addition, the calculations in the $f$-model space (including the $f_{7/2}$ 
and the $f_{5/2}$ orbits only) using the same Hamiltonian are given in 
Fig.~\ref{DGTfspace}. For $^{42}$Ca, there are two $0^+$ DGT states at the 
excitation energies $0.0$ MeV and $18.3$ MeV. Their strengths are $11.178$ and 
$13.791$, respectively. There is one $2^+$ DGT state at $0.0$ MeV and its 
strength is $8.658$. Note that the sum rules do not depend on the model space. 
In the $f$-model space, we obtained exactly the sum rules even for the case of 
the DGT to $2^+$ in $^{48}$Ca because the calculation can be done without 
any limitation as now the total number of possible final states is strongly 
reduced (see Table \ref{SDGT}). The DGT distributions in the $f$-model space are 
much more concentrated. The analytical calculation in the limited $f$-model 
space helps us know where the DGT strengths concentrate.

\section{Conclusions}
The general features and trends of the DGT sum rules in even-$A$ Calcium 
isotopes are described using the numerical results. The properties of the 
entire distributions of the DGT are discussed. 
By studying the stronger DGT transitions, in particular, the DGT giant 
resonance experimentally and theoretically, the calculations of 
$\beta\beta$-decay nuclear matrix elements can be calibrated to some extent. 
There is no doubt that the pion DCX is a sensitive probing tool 
of nuclear structure. Nowadays the ion DCX reactions have been discussed 
mainly in the context of $0\nu\beta\beta$, however, the ion DCX reaction 
itself is a new probing tool of nuclear structure, in particular of spin 
degrees of freedom. The DGT resonance is just one example. Because two 
nucleons participate in the DCX reactions with pions or heavy ions, one can 
expect that the nucleon-nucleon interaction and correlations can be probed, 
especially for the nucleus that is far from the stability regime.

\begin{acknowledgments}
The authors thank to B. A. Brown and Vladimir Zelevinsky for discussions made 
possible by the travel grant from the US-Israel Binational Science Foundation 
(2014.24). This work was supported by the US-Israel Binational Science 
Foundation, grant 2014.24.
\end{acknowledgments}

\end{document}